\documentclass[fleqn,usenatbib]{mnras}

\usepackage{newtxtext,newtxmath}
\usepackage[T1]{fontenc}
\usepackage{ae,aecompl}
\usepackage{amsfonts, graphicx, amsmath, amssymb}
\usepackage{nccmath}
\usepackage{hyperref}

\title[Probing CMB Lensing with Quasars]{Probing Gravitational Lensing of the CMB with SDSS-IV Quasars}

\author[J. Han et al.]{
Jiashu Han $^{1}$\thanks{E-mail: jiashu.han@berkeley.edu},
Simone Ferraro$^{2,3}$,
Elena Giusarma$^{4,5}$
and Shirley Ho $^{2,3,4}$
\\
$^{1}$Department of Physics, University of California, Berkeley, CA 94720, U.S.A.\\
$^{2}$Lawrence Berkeley National Laboratory, 1 Cyclotron Rd, Berkeley, CA 94720, U.S.A.\\
$^{3}$Berkeley Center for Cosmological Physics, University of California Berkeley, Berkeley, CA 94720, U.S.A.\\
$^{4}$Center for Computational Astrophysics, Flatiron Institute, 162 5th Avenue, New York, NY 10010, U.S.A.\\
$^{5}$McWilliams Center for Cosmology, Department of Physics, Carnegie Mellon University, Pittsburgh, PA 15213, USA
}

\begin{document}
\label{firstpage}
\pagerange{\pageref{firstpage}--\pageref{lastpage}}
\maketitle

\begin{abstract}
We study the cross-correlation between the \textit{Planck} CMB lensing convergence map and the eBOSS quasar overdensity obtained from the Sloan Digital Sky Survey (SDSS) IV, in the redshift range $0.9 < z < 2.2$. We detect the CMB lensing convergence-quasar cross power spectrum at $5.4 \sigma$ significance. The cross power spectrum provides a quasar clustering bias measurement that is expected to be particularly robust against systematic effects. The redshift distribution of the quasar sample has a median redshift $z \approx 1.55$, and an effective redshift about $1.51$. The best fit bias of the quasar sample is $b_q = 2.43 \pm 0.45$, corresponding to a host halo mass of $\log_{10}\left( \frac{M}{h^{-1} M_\odot} \right) = 12.54^{+0.25}_{-0.36}$. This is broadly consistent with the previous literature on quasars with a similar redshift range and selection. Since our constraint on the bias comes from the cross-correlation between quasars and CMB lensing, we expect it to be robust to a wide range of possible systematic effects that may contaminate the auto correlation of quasars. We checked for a number of systematic effects from both CMB lensing and quasar overdensity, and found that all systematics are consistent with null within $2 \sigma$. The data is not sensitive to a possible scale dependence of the bias at present, but we expect that as the number of quasars increases (in future surveys such as DESI), it is likely that strong constraints on the scale dependence of the bias can be obtained.
\end{abstract}



\section{Introduction}
Cosmic microwave background (CMB) temperature fluctuations provide invaluable information about our Universe, and can give extremely tight constraints on  cosmological parameters \citep{Kofman1993, Hinshaw2012, Planck2015-params, Planck2018-params}. The primary CMB anisotropy encodes information about the primordial universe, measured at $z \approx 1100$. However, since the discovery of the CMB, a lot of progress has also been made on the secondary CMB anisotropies, such as gravitational lensing \citep{Smith2007, Hirata2008, Lewis2006}, the thermal and kinetic Sunyaev-Zel'dovich (tSZ, kSZ) effects \citep{Sunyaev1980}, and the integrated Sachs-Wolfe effect (ISW) \citep{Sachs1967}. These effects can act as foreground for the primary CMB, but they also encode information about the growth of structure at lower redshifts, a powerful probe of Dark Energy, Modified Gravity and neutrino masses \citep{Lewis2006}. 


Secondary CMB anisotropies are produced by large scale structures (LSS) in the late-time universe \citep{Aghanim2008} and can be easily detected through cross-correlation with the LSS. In particular, CMB lensing traces the matter density field at intermediate redshifts ($z \lesssim 1100$). As CMB photons travel to the observer, they are gravitationally deflected by the matter, leaving an imprint on the observed CMB temperature and polarization fluctuations. Weak lensing of the CMB introduces off-diagonal correlations between Fourier modes, allowing the CMB lensing deflection field $\bf{d}$ to be estimated \citep{Hu2002}. 

We use the CMB lensing convergence field as a tracer of the dark matter field, and thus the large scale structure of the Universe. More specifically, we use the CMB lensing convergence field to study properties of quasars, which trace the large scale structure at intermediate redshifts. Quasars are thought to be luminous accreting supermassive black holes at the centers of distant galaxies \citep{Salpeter1964}. Like galaxies, they are tracers of the 3D distribution of dark matter at different redshifts. With the understanding that almost every galaxy hosts a supermassive black hole at its center \citep{Kormendy1995}, quasars can be thought as a phase in the galaxy evolution. Properties of quasars, such as the characteristic mass of the their host halos \citep{Tinker2010, DiPompeo2014}, can be inferred by studying the relationship between the dark matter distribution and quasar clustering. The information about quasar properties can reveal much about the growth of structure over the history of the universe \citep{Marziani2014, Mortlock2015}. 

Both CMB lensing and the observed quasar overdensity depend on the projected matter overdensity, and the quasar redshift distribution matches well with the CMB lensing kernel, so the CMB lensing convergence and the quasar overdensity should have a relatively strong correlation \citep{Peiris2000}. Quasars are biased tracers of the underlying matter density field \citep{Kaiser1984}, meaning that the observed cross power spectrum is proportional to the quasar bias. This factor parametrizes the properties of the clustering of quasars and encapsulates the information about the processes of galaxy formation and evolution that are currently not very well understood \citep{Amendola2017}. Measuring this bias factor would be crucial to the understanding of galaxy formation and the evolution of supermassive black holes within the standard structural formation framework \citep{Shen2009}.

\citet{Laurent2017} have analyzed the auto-correlation of the eBOSS quasars, and put constraints on the quasar bias, as well as the corresponding host halo mass. In this paper, we use an alternative way to constrain the quasar bias, by cross-correlating the CMB lensing map from Planck and quasar overdensities drawn from eBOSS Data Release 14 \citep{Dawson2016}. We measure a quasar bias that is consistent with the auto-correlation result. We then calculate the corresponding characteristic host halo mass of the quasars. All calculations assume a cosmology with the Planck TT+lowP+lensing parameters \citep{Planck2015-params}. 

The remainder of the paper is organized as follows: in Section \ref{sec:background} we present the theoretical background of quasar linear bias and the CMB lensing-quasar angular cross power spectrum. Section \ref{sec:methods} includes the data samples and estimators we used to evaluate the observed power spectrum. In Section \ref{sec:results}, we estimate the cross power spectrum, the quasar bias, and the characteristic host halo mass. In Section \ref{sec:error}, we discuss errors, systematic effects and a null test performed on the data. We draw our conclusions in Section \ref{sec:conclusions}.

\begin{figure}
	\includegraphics[width=\columnwidth]{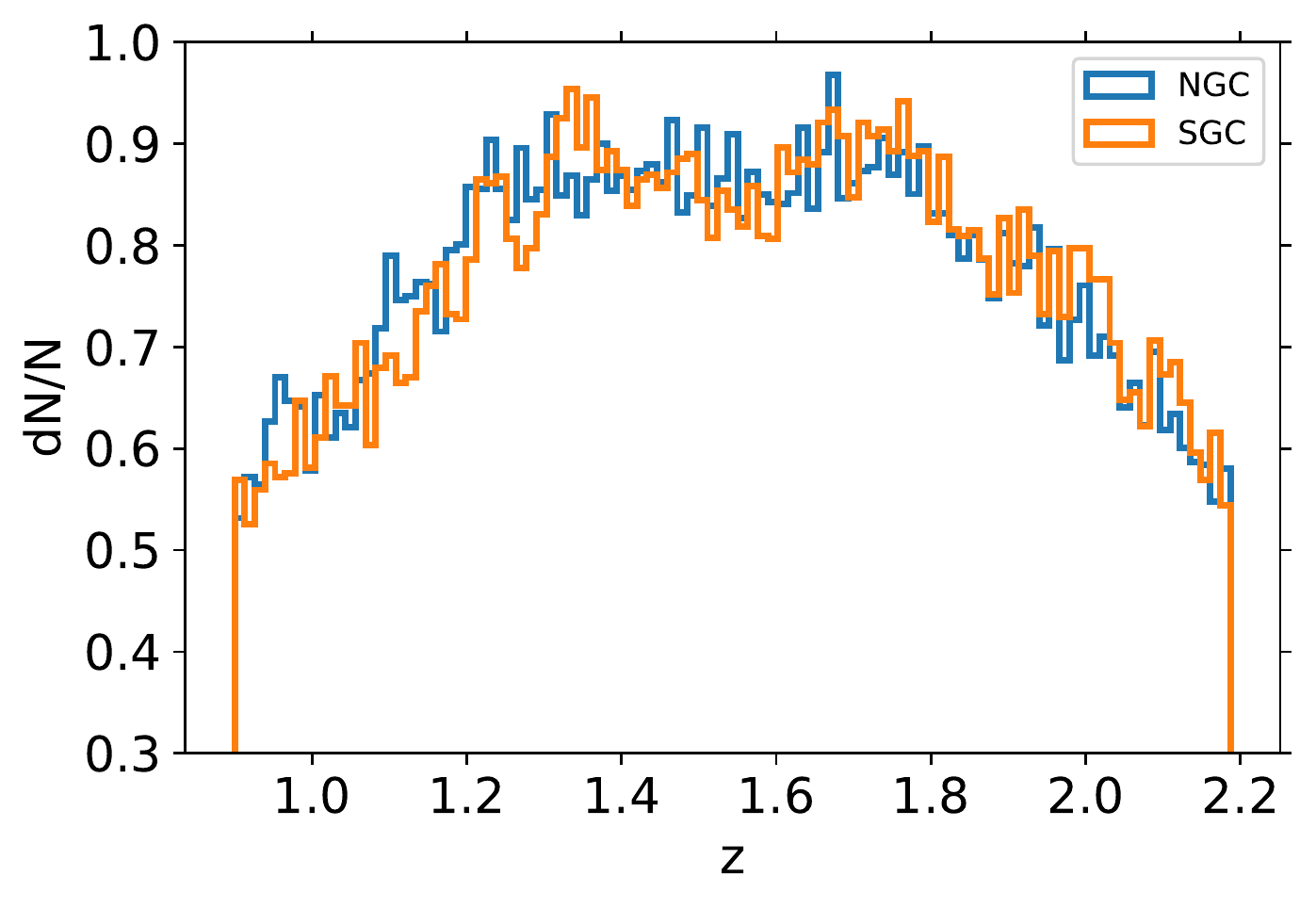}
    \caption{The redshift distributions of the selected quasars in the range 0.9 < $z$ < 2.2 in the North and South Galactic Caps (NGC, in blue; and SGC, in orange). The normalized redshift distribution is plotted on the y-axis.}
    \label{fig:fz}
\end{figure}

\section{Theoretical Background}
\label{sec:background}
\subsection{Overview}
Quasars reside in the nuclei of distant galaxies and are expected to be biased tracers of the matter overdensity on large scales. In other words, the number density of quasars is related to the dark matter overdensity by a bias factor, i.e. $\delta_q = b_q \ \delta_m$, where $b_q$ can be a function of scale, redshift, formation history or other environment related factors \citep{White1978}. The amplitude of the deflection by CMB lensing in a given direction depends on the projected matter density in that direction. Thus we expect the quasar number density to be correlated with CMB lensing convergence \citep{Peiris2000}. 


\subsection{Angular cross power spectrum}
\label{sec:powspec}
To relate the CMB lensing to the matter overdensity field, we define the lensing convergence, $\kappa \equiv -\frac{1}{2} \nabla \cdot \bf{d}$, where $\bf{d}$ is the lensing deflection field. The lensing convergence is a weighted projection of the matter overdensity in direction $\hat{n}$ along the line of sight \citep{Lewis2006}:
\begin{ceqn}
\begin{equation}
\label{eq:kappa}
\kappa(\hat{n}) = \int_0^{z_{\textrm{\tiny CMB}}} dz W(z) \delta_m(\chi(z) \hat{n}, z)
\end{equation}
\end{ceqn}
where $z_{\textrm{\tiny CMB}} \approx 1100$ is the redshift at the last scattering surface, $W(z)$ is the CMB lensing kernel, $\delta_m(\chi(z) \hat{n}, z)$ is the matter overdensity at redshift $z$ in the direction $\hat{n}$, and $\chi(z)$ is the comoving distance at redshift $z$. Assuming a flat universe, $W(z)$ is given by
\begin{ceqn}
\begin{equation}
\label{eq:wz}
W(z)= \frac{3 H_0^2 \Omega_{m,0}}{2 c H(z)} (1+z) \chi(z)\left( 1 - \frac{\chi(z)}{\chi_{\textrm{\tiny CMB}}}\right)
\end{equation}
\end{ceqn}
where $\chi_{\textrm{\tiny CMB}}$ is the comoving distance to the last scattering surface, $H_0$ is the current Hubble parameter, $H(z)$ is the Hubble parameter at redshift $z$, and $\Omega_{m, 0}$ is the current matter density parameter. Since the lensing potential $\phi$ is a 2D projection of the gravitational potential, we can assume CMB lensing as an unbiased tracer of the underlying matter overdensity field \citep{Lewis2006}.

The quasar overdensity field is related to the matter overdensity field by a window function $f(z)$, such that the projected surface density is $q(\hat{n}) = \int_0^{z_{\textrm{\tiny CMB}}} dz f(z) \delta_m(\chi(z)\hat{n}, z)$ \citep{Peiris2000}:
\begin{ceqn}
\begin{equation}
\label{eq:fz}
f(z)= \frac{b(z) dN/dz}{\int dz' \frac{dN}{dz'}} + \frac{3}{2 H(z)} \Omega_0 H_0^2 (1+z) g(z) (5s - 2).
\end{equation}
\end{ceqn}
In the previous equation, the first term is the normalized, bias-weighted redshift distribution of the quasars. The second term is the magnification bias, which accounts for the change in the density of the sources due to lensing magnification \citep{Moessner1997, Scranton2005}. This term is negligible compared to the intrinsic clustering of the quasars, and for this reason we ignore it for simplicity. For a full expression of $g(z)$, see \citet{Peiris2000, Sherwin2012}.

On large scales, we expect the quasar bias to be a constant. On smaller scales, however, the scale-dependence of the bias has been supported by many measurements and it is predicted by theory \citep{Amendola2017, Giusarma2018}. 

Many bias models have been proposed. We will consider an effective power law parametrization of the scale dependence of the bias:
\begin{ceqn}
\begin{equation}
\label{eq:scale-de}
	b (k) = b_1 + b_2 \left( \frac{k}{k_0} \right)^n
\end{equation}
\end{ceqn}
where $k_0$ is an arbitrary reference scale we set to be $1 h \textrm{ Mpc}^{-1}$, such that $b_2$ is a dimensionless parameter \citep{Amendola2017}. The case $n = 0$ corresponds to a scale-independent bias.

\citet{Desjacques2016} and \citet{Modi2017} reported an n = 2 behavior at scales $0.1 \lesssim k \lesssim 0.5h~\textrm{Mpc}^{-1}$ for the linear halo bias, based on results from N-body simulations. We will test this form for the scale-dependent quasar bias. 

If the selection functions of the dark matter tracers are slowly varying compared to the scale we are probing, the Limber approximation \citep{Limber1954, Lewis2006} is expected to be valid at $\ell \gtrsim 30$. Assuming a flat universe, the quasar-CMB lensing convergence angular cross-power spectrum is given by:
\begin{ceqn}
\begin{equation}
\label{eq:cl}
	C_l^{\kappa q} = \int \frac{dz}{c} \frac{H(z)}{\chi^2(z)} W(z) f(z) P_{mm}\left(k=\frac{l}{\chi(z)}, z\right)
\end{equation}
\end{ceqn}
where $f(z)$ is the bias-weighted redshift distribution, and $P_{mm}(k, z)$ is the 3D matter power spectrum. 

An advantage of using the cross-correlation between CMB lensing and quasars over doing quasar auto-correlation, is that the quasar clustering-matter cross power spectrum has a linear dependence on the quasar bias, from the bias-weighted redshift distribution. Moreover, measuring this cross-correlation in addition to the auto-correlation of quasars helps break the degeneracy between quasar bias $b_q$ and amplitude of fluctuations\footnote{This is because the auto-correlation measures $b_q^2 \sigma_8^2$, while the cross correlation is proportional to $b_q \sigma_8^2$} $\sigma_8$, thus improving our constraints on $\sigma_8$ . The cross-correlation is also less likely to be affected by systematics in the quasar sample \citep{Sherwin2012, Geach2013, 2012ApJ...753L...9B}.

\begin{figure}
	\includegraphics[width=\columnwidth]{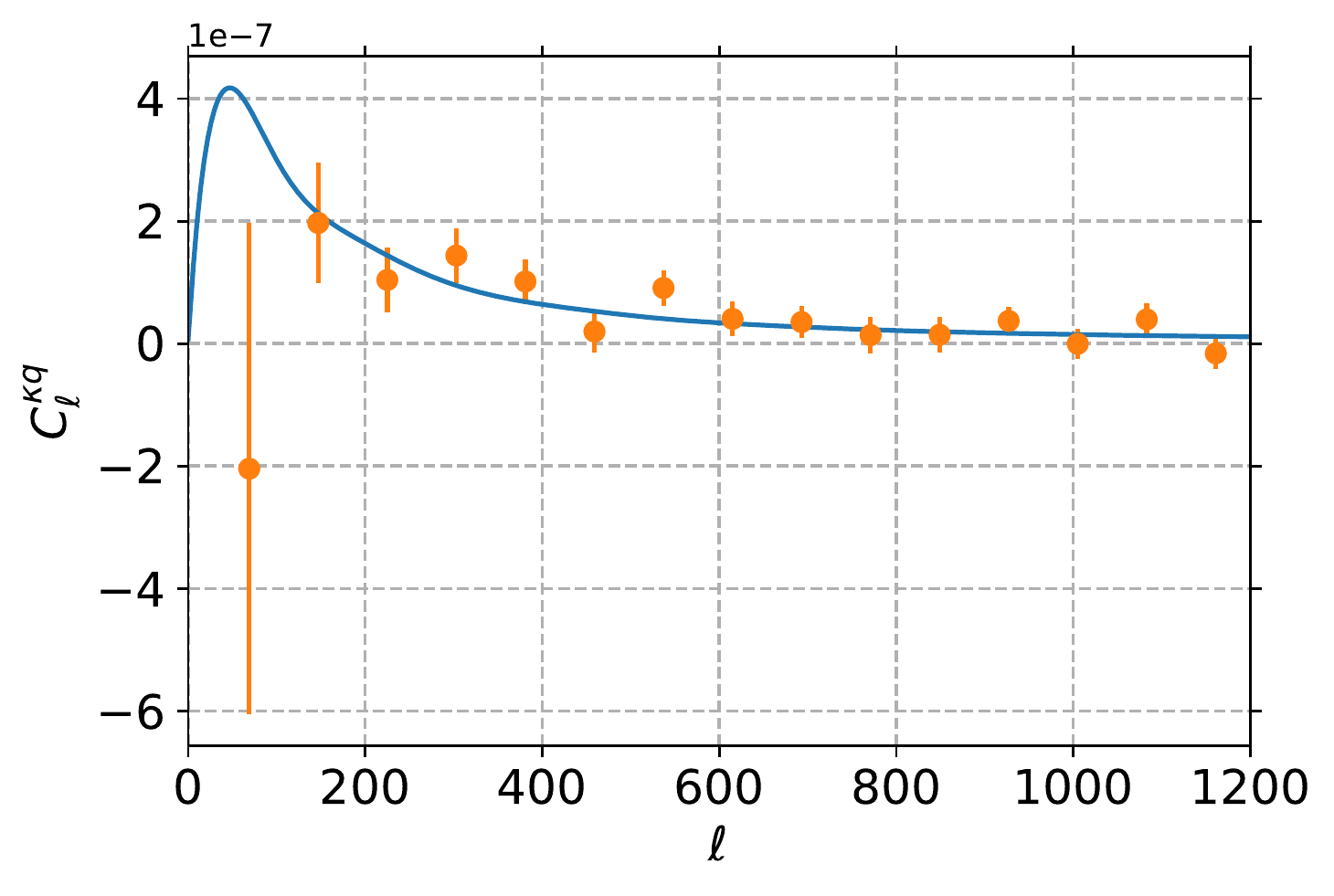}
    \caption{The CMB lensing-quasar overdensity angular cross-power spectrum. The data points are in orange, and the blue solid curve is the calculated theory curve. The significance of the cross-power spectrum signal is 5.4$\sigma$.}
    \label{fig:cl}
\end{figure}

\section{Data and Methods}
\label{sec:methods}

\subsection{CMB lensing map}
We use the CMB lensing convergence map published by the Planck Collaboration \citep{Planck2015-lensing}. The \textit{Planck} satellite, which was launched in 2009, observed the temperature and polarization fields of the cosmic background radiation over the whole sky at various frequencies.
Maps of the temperature and polarization fields of the CMB covering 70\% of the sky are produced \citep{Planck2015-overview}. The Planck minimum-variance CMB lensing potential field is reconstructed using the CMB maps produced by the SMICA code, and combines the five quadratic estimators of the correlations of the CMB temperature ($T$) and polarizations ($E, B$). The map underwent several systematic and null tests, that showed that any contamination is small compared to the statistical errors.

\subsection{Quasar map}
We use quasars from the extended-Baryon Oscillation Spectroscopic Survey (eBOSS, \citet{Dawson2016, Zhao2016}), which started in July 2014, as an extension to the Baryon Oscillation Spectroscopic Survey (BOSS) \citep{Dawson2013}. BOSS probed the BAO at a scale of roughly 100 $h^{-1}$ Mpc, using mostly galaxies at $z < 0.7$ and neutral hydrogen clouds in the Lyman-$\alpha$ forest at $z > 2.1$. 

eBOSS aims to probe four different dark matter tracers at redshift ranges that are not covered in previous surveys, and map the large scale structures over the redshift range $0.6 < z < 2.2$, which is previously unconstrained by BOSS. The full eBOSS quasar catalog \citep{Myers2015} is expected to contain 500,000 spectroscopically-confirmed quasars over an area of 7500 deg$^2$ by the end of the survey and provide the first BAO distance measurement over the range $0.9 < z < 2.2$. The eBOSS quasars will also provide tests of General Relativity on the cosmological scales through measurements of the redshift-space distortion, and new constraints on the summed mass of all known neutrino species.

We use the quasars from the eBOSS DR14 LSS catalog\footnote{\url{https://data.sdss.org/sas/ebosswork/eboss/lss/catalogs/catalogs-DR14/}} \citep{Myers2015}, which contains 142,017 quasars between $0.9 < z < 2.2$ and has an effective redshift of 1.51. The redshift distribution of the selected quasars is shown in Fig.~\ref{fig:fz}. We construct an overdensity map ($q_i = \frac{n_i - \bar{n}}{\bar{n}}$, where $i$ is the pixel number) of these quasars. The map is converted into HEALPix format with $N_{\textrm{side}} = 2048$ to match the resolution of the CMB lensing convergence map. We find the quasar footprint by downgrading the resolution of the quasar map to $N_{\textrm{side}} = 32$ and identifying the empty pixels in the map.


\begin{figure}
	\includegraphics[width=\columnwidth]{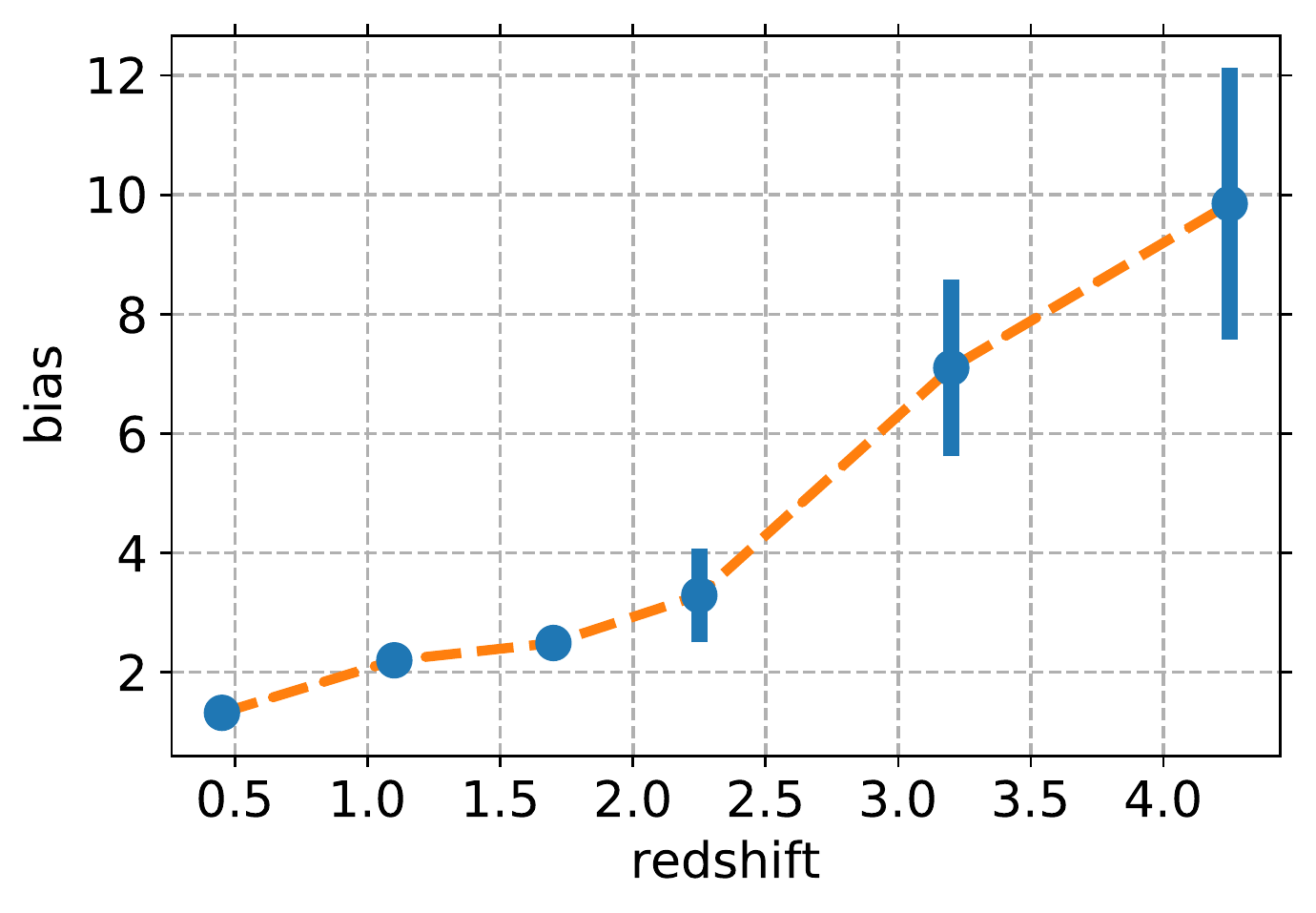}
    \caption{The fiducial bias-redshift model used in the calculation, obtained by interpolating the data points in \citet{Shen2009}. The paper also provides estimates of the error in the quasar bias, which are shown as error bars in the plot. The dashed line in orange is the interpolated result.}
    \label{fig:fid}
\end{figure}

\subsection{Estimator for the angular power spectrum}
We use a pseudo-$C_l$ estimator \citep{Wandelt2000} to calculate the angular cross-power spectrum from the data
\begin{ceqn}
\begin{equation}
\hat{C}_l^{\kappa q}=\frac{1}{f_{\textrm{sky}}^{\kappa q}(2l + 1)}\sum_{m=-l}^{l}{\kappa_{lm}^* q_{lm}}
\end{equation}
\end{ceqn}
where $f_{\textrm{sky}}^{\kappa q}$ is the fraction of the sky shared by the quasar map and the CMB lensing convergence map. $\kappa_{lm}$ is the spherical harmonic transform of the CMB lensing convergence map, and $q_{lm}$ is the spherical harmonic transform of the quasar overdensity map. 

In the Fisher approximation, the theoretical error in each bin $A$ of $\hat{C_l}^{\kappa q}$ can be estimated using \citep{Cabre2008}
\begin{ceqn}
\begin{equation}
\label{eq:error}
\frac{1}{\sigma^2(A)}= \sum_{l_{\textrm{min}}(A) < l < l_{\textrm{max}}(A)} \frac{f_{\textrm{sky}}^{\kappa q}(2l + 1)}{(C_l^{\kappa q})^2+C_l^{\kappa \kappa} C_l^{q q}}
\end{equation}
\end{ceqn}
where $C_l^{\kappa \kappa}$ and $C_l^{q q}$ are the CMB lensing and quasar auto-power spectra, including both signal and noise. The contribution of error from the $C_l^{\kappa \kappa} C_l^{qq}$ term should dominate the contribution from the cross term. The  auto-spectra can be estimated similarly:
\begin{ceqn}
\begin{equation}
\hat{C}_l^{\kappa \kappa}=\frac{1}{f_{\textrm{sky}}^{\kappa}(2l + 1)}\sum_{m=-l}^{l}{|\kappa_{lm}|^2}
\end{equation}
\end{ceqn}
and
\begin{ceqn}
\begin{equation}
\hat{C}_l^{q q}=\frac{1}{f_{\textrm{sky}}^{q}(2l + 1)}\sum_{m=-l}^{l}{|q_{lm}|^2}
\end{equation}
\end{ceqn}
where $f_{\textrm{sky}}^{\kappa}$ is the sky fraction of the CMB lensing convergence map, and $f_{\textrm{sky}}^{q}$ is the sky fraction of the quasar overdensity map.  We bin the cross-power spectrum into 15 bands in the range $30 \le \ell < 1200$. We choose $\ell_{\textrm{min}} = 30$ because the Limber approximation breaks down on larger scales. We choose $\ell_{\textrm{max}} = 1200$ because of the uncertainty on modeling the bias and power spectrum on smaller scales. The signal-to-noise also drops significantly for $\ell > \ell_\textrm{max}$ \citep{Lewis2006,Kirk2015}.

\section{Results}
\label{sec:results}
\subsection{Cross-correlation}
The cross-correlation results are shown in Fig.~\ref{fig:cl}. The theoretical curve is calculated using Equation~\ref{eq:cl}. We use the redshift distribution in Fig.~\ref{fig:fz} and the CMB lensing kernel in Equation~\ref{eq:wz}. The sample variance fluctuations are of order $\sim10\%$ per bin in redshift, due to the large number of quasars in the bin. We use the full $dN/dz$ from Figure \ref{fig:fz} in the theory calculation, which yields a smooth result since it is integrated over a broad kernel. The theory curve should not be sensitive to binning and interpolation, since the weighting functions are slowly varying with redshift. We use CAMB \citep{Lewis2000} to compute the matter power spectrum. The nonlinear matter power spectrum (HALOFIT, \citet{Smith2003, Takahashi2012}) is used in this calculation. The linear matter power spectrum produces similar results because the signal mainly comes from angular scales ($\ell < 600$) corresponding to linear scales.

We assume a fiducial bias-redshift model from \citet{Shen2009} in the theory calculation, shown in Fig.~\ref{fig:fid}. The fiducial bias model is based on the amplitude of the quasar correlation function from the SDSS DR5 quasar sample. We fit a scaled version of the fiducial bias-redshift relation to the data to find the best-fit scaling parameter ($b_q/b_\textrm{fid}$), and the theory curve is a good fit. With 14 degrees of freedom, the chi-squared value for the best-fit theory curve is $\chi^2_{\textrm{th}} = 12.9$. The significance of the cross-correlation is $\sqrt{\chi^2_0 - \chi^2_{\textrm{th}}} = 5.4\sigma$, where $\chi^2_0$ is the chi-squared value for the null hypothesis. The detection significance is also the best-fit scaling parameter divided by its error.

All points are included in the model fits to the data. Despite the theory curve being a good fit to the data points, the first bin is more than $1 \sigma$ from the theoretical prediction and shows an anti-correlation between the CMB lensing map and the quasar overdensity, albeit having a large uncertainty. \citet{Giannantonio2016}, which uses the CMB lensing data from the South Pole Telescope \citep{Story2015}, and \citet{Pullen2016} also reported a deficit of power in the low $\ell$ region of the CMB lensing-galaxy angular cross power spectrum. We do not have an explanation for the cause of this deficit of power. However, we can rule out a list of systematics, described in detail later in Section \ref{sec:error}, as causes of this deficit.

\subsection{Quasar bias}
The fiducial bias-redshift model used in the calculation is obtained by interpolating the data in \citet{Shen2009}. Although it uses a different quasar catalog than the one in our analysis, we choose this as a convenient model because the theoretical cross-power spectrum does not have a strong dependence on the detailed form of the bias model \citep{Sherwin2012}. The fiducial model is shown in Fig.~\ref{fig:fid}. From this model we find $b_q/b_{\textrm{fid}} = 1.01 \pm 0.19$. At the effective redshift of our quasar sample ($z \approx 1.51$), the fiducial model gives a bias $b_{\textrm{fid}} = 2.4$. Combining these results we get a quasar linear bias of $b_q = 2.43 \pm 0.45$, with $5.4 \sigma$ significance.

We also fit for the scale-dependent bias in Equation~\ref{eq:scale-de}, by fixing $n$ at various values. Table~\ref{tab:scale-dep} shows some of the results. In the $n = 2$ case, we have $b_1 = 2.26 \pm 0.59$ and $b_2 = 13.0 \pm 33.3$. We conclude that the data does not yield a strong constraint on the scale dependent bias. This, however, could be due to the low number density of quasars in the survey. We expect that the better sensitivity and resolution of future surveys will allow better constraints on the scale dependence of both the bias and the matter power spectrum \citep{Abazajian2016}.

\begin{table}
\centering
\begin{tabular}{lccccr}\hline
$n$ & $b_0$ & $\sigma(b_0)$ & $b_1$ & $\sigma(b_1)$ & $\chi^2$ \\\hline
-2 & 2.80 & 0.50 & -0.0010 & 0.0006 & 11.0 \\
-1 & 3.53 & 0.81 & -0.067 & 0.041 & 11.2 \\
0 & 2.43 & 0.45 & - & - & 12.9 \\
1 & 1.85 & 0.89 & 6.37 & 8.71 & 13.3 \\
2 & 2.26 & 0.59 & 13.0 & 33.3 & 13.7 \\
\end{tabular} 
\caption{\label{tab:scale-dep}Selected results for the scale-dependent bias fit. In the third row, $n = 0$ corresponds to a scale-independent bias.}
\end{table}

\subsection{Quasar host halo mass}
As shown in Fig.~\ref{fig:fid}, the quasar bias generally increases with redshift, and the bias is expected to increase with halo mass. However, at higher redshifts, the halos also have less time to grow. Therefore, we would expect a roughly constant halo mass-redshift relation.

We use the bias model provided in \citet{Tinker2010} to relate the scale-independent quasar bias to the peak height of the linear density field, $\nu = \frac{\delta_c}{\sigma(M)}$, where $\delta_c = 1.686$ is the critical overdensity for collapse, and calculate a corresponding characteristic host halo mass. We assume the ratio between the halo mass density and the average matter density of universe is $\Delta = 200$. We find the characteristic host halo mass to be $\log_{10}\left( \frac{M}{ h^{-1} M_\odot} \right) = 12.54^{+0.25}_{-0.36}$. This is consistent with previous estimates for BOSS/eBOSS quasars at similar redshifts \citep{White2012, Laurent2017}.

\begin{figure}
	\includegraphics[width=\columnwidth]{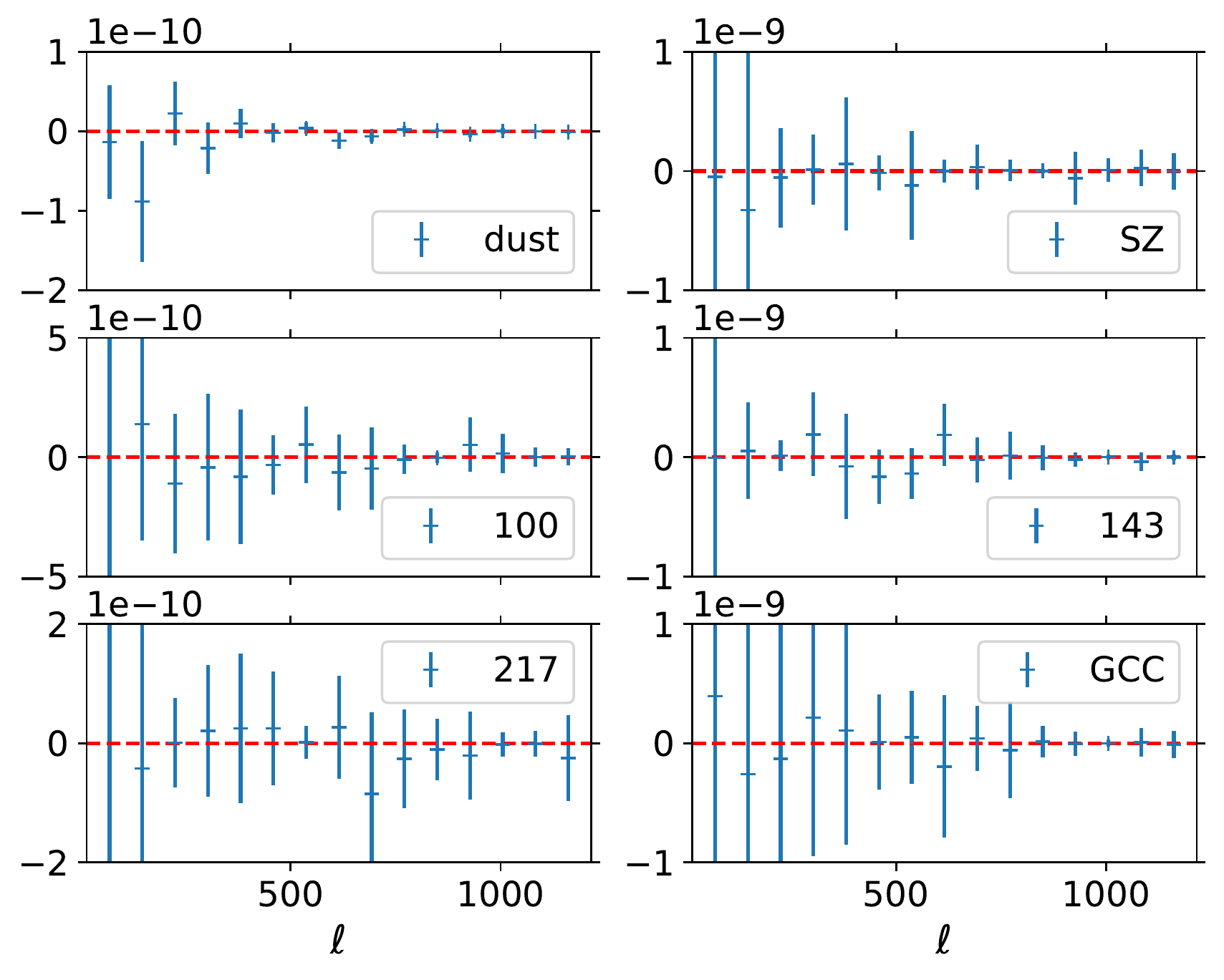}
    \caption{Check for possible systematic effects on the cross power spectrum due to contaminants. Here we show the right hand side of Equation \ref{eq:systematics} for different foregrounds. The dust plot is the bias due to dust emission. The SZ plot is the bias due to the Planck SZ catalog. The 100, 143, and 217 plots are the biases from Planck Catalog of Compact Objects, corresponding to the labeled frequency. The GCC plot is the bias from the Planck Galactic cold clumps.}
    \label{fig:sys}
\end{figure}

\section{Measurement systematics and uncertainties}
\label{sec:error}
\subsection{Systematic effects}
Residual foregrounds in the CMB map that are correlated with the large scale structure probed by the eBOSS quasars can lead to biases to the CMB lensing cross-correlation \citep{2014ApJ...786...13V, 2014JCAP...03..024O, 2018PhRvD..97b3512F,Pullen2016}. Mitigation strategies have been proposed \citep{2018arXiv180208230M, 2018arXiv180406403S}, and based on previous work we expect the bias to cross-correlations with Planck lensing to be at most a few percent, considerably smaller than our statistical significance.

Nonetheless, we check for contamination from galactic dust emission, point sources, and SZ effect. We use the Second Planck SZ Catalog \citep{Planck2015-sz}, which includes sources detected through the SZ effect \citep{Sunyaev1980}, the \citet{Schlegel1998} dust infrared emission map for estimation of CMB radiation foregrounds, the \textit{Planck} Catalog of Galactic cold clumps \citep{Planck2015-gcc}, and the overdensity maps constructed from the Second \textit{Planck} Catalog of Compact Sources \citep{Planck2015-ccs} at frequencies 100 GHz, 143 GHz, and 217 GHz.

If systematic effects were added linearly to the observed CMB lensing map and quasar map \citep{Ross2011, Ho2012}, the bias to the cross correlation would be given by \citep{Giannantonio2016}:
\begin{ceqn}
\begin{equation}
\label{eq:systematics}
\Delta \hat{C}_{l}^{\kappa q} = \sum_s \frac{\hat{C}_{l}^{\kappa s} \hat{C}_{l}^{q s}}{\hat{C}_{l}^{s s}}
\end{equation}
\end{ceqn}
where $s$ is the map for the systematics. In Equation \ref{eq:systematics}, we estimate the amplitude of the systematic $s$ for each data set by cross-correlating the data and the systematic template, and propagate these to the bias in the observed cross power spectrum. Although the lensing map is obtained though non-linear operations on the CMB map, and therefore the assumption of linearity is not satisfied, estimating the quantity above is still a powerful null-test. If significant contamination was found, Equation \ref{eq:systematics} should not be used to correct for the bias, but more sophisticated mitigation techniques should be employed \citep{2018arXiv180406403S, 2018arXiv180208230M, 2014JCAP...03..024O}.

Fig.~\ref{fig:sys} shows the right hand side of Equation \ref{eq:systematics}. The effects are consistent with null at most scales and we conclude that there is no significant systematic effects due to the contaminants considered above. We calculate the overall systematic error by adding the average absolute biases at each angular scale, weighted by the inverse variance, and find it to be less than 7\% of the signal.

\subsection{Null test}
\begin{figure}
	\includegraphics[width=\columnwidth]{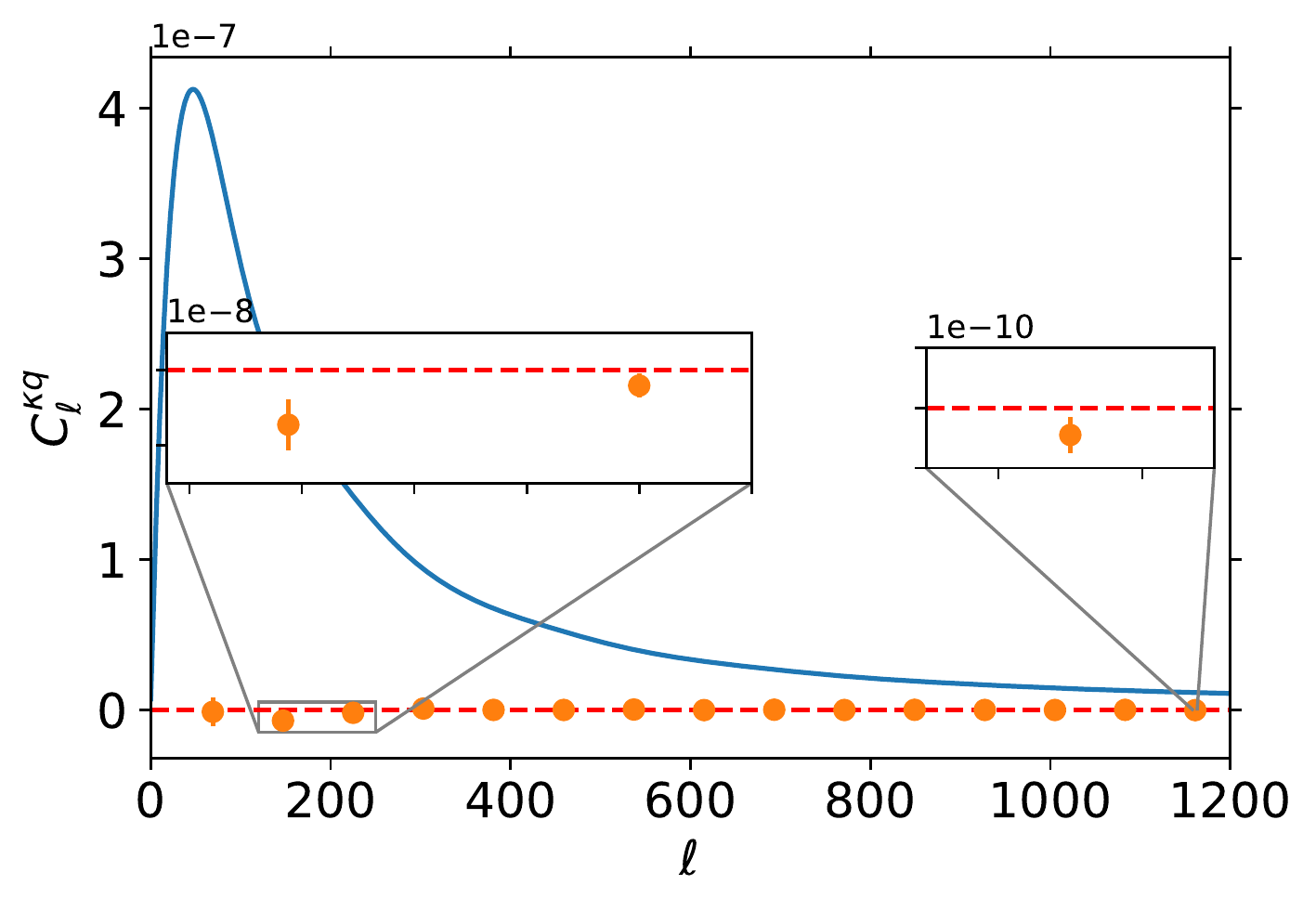}
    \caption{The cross-power spectrum from the null test. The error bars are obtained the same way as before (Equation \ref{eq:error}). The blue curve is the best-fit theoretical cross-power spectra. The zoomed-up subplots show points more than $1 \sigma$ from null. The result is consistent with zero correlation.}
    \label{fig:null}
\end{figure}

We use a simple null test \citep{Sherwin2012, Geach2013} on the CMB lensing-quasar overdensity cross power spectrum to check our result and procedure by cross-correlating the CMB lensing convergence map on one part of the sky with the quasar map on another part of the sky. The result of the null test is shown in Fig.~\ref{fig:null}. Most bins of the null cross-spectrum fall within $1\sigma$ of null, and fitting the theoretical spectrum to the null result yields a bias measurement of $b / b_\textrm{fid} = -0.005 \pm 0.003$. The best fit chi-square value for the null hypothesis is 11.23, with 14 degrees of freedom. The distribution of the points is consistent with a Gaussian centered at zero.

\subsection{Covariance matrix}
\label{cov-matrix}
The theoretical error bar for each bin is calculated using Equation~\ref{eq:error}, which assumes the bins are independent. Limited sky fraction may induce correlation between $\ell$ bins, and this assumption is only valid when the bins are relatively large ($\Delta\ell \gtrsim 2/f_{\textrm{sky}} $) \citep{Gaztanaga2012, Cabre2008}. In our case, the bins are large and should be roughly independent in the limit of large $\Delta \ell$.

To compute the full covariance matrix of $\hat{C}_l^{\kappa q}$, we use quasar mocks and CMB lensing simulations. The quasar mocks are taken from the QSO EZmocks (effective Zel'dovich approximation mock catalogs) \citep{Chuang2014}, which include 1000 realizations of the quasar map with the same number of randomly distributed sources. The CMB lensing simulations include 100 realizations of simulated lensing convergence maps \citep{Planck2015-lensing} containing both signal and noise. We cross correlate 100 pairs of the quasar mocks and lensing simulations, and calculate the average of the cross power spectra, to estimate the covariance matrix $\textrm{cov}[i, j] = \langle (C(i) - E(C(i)))(C(j) - E(C(j))) \rangle$. Note that the covariance estimated via this route does not include the $C_l^{\kappa q}$ part in Equation \ref{eq:error}, because the quasar mocks are not correlated with the CMB lensing simulations.

\begin{figure}
	\includegraphics[width=\columnwidth]{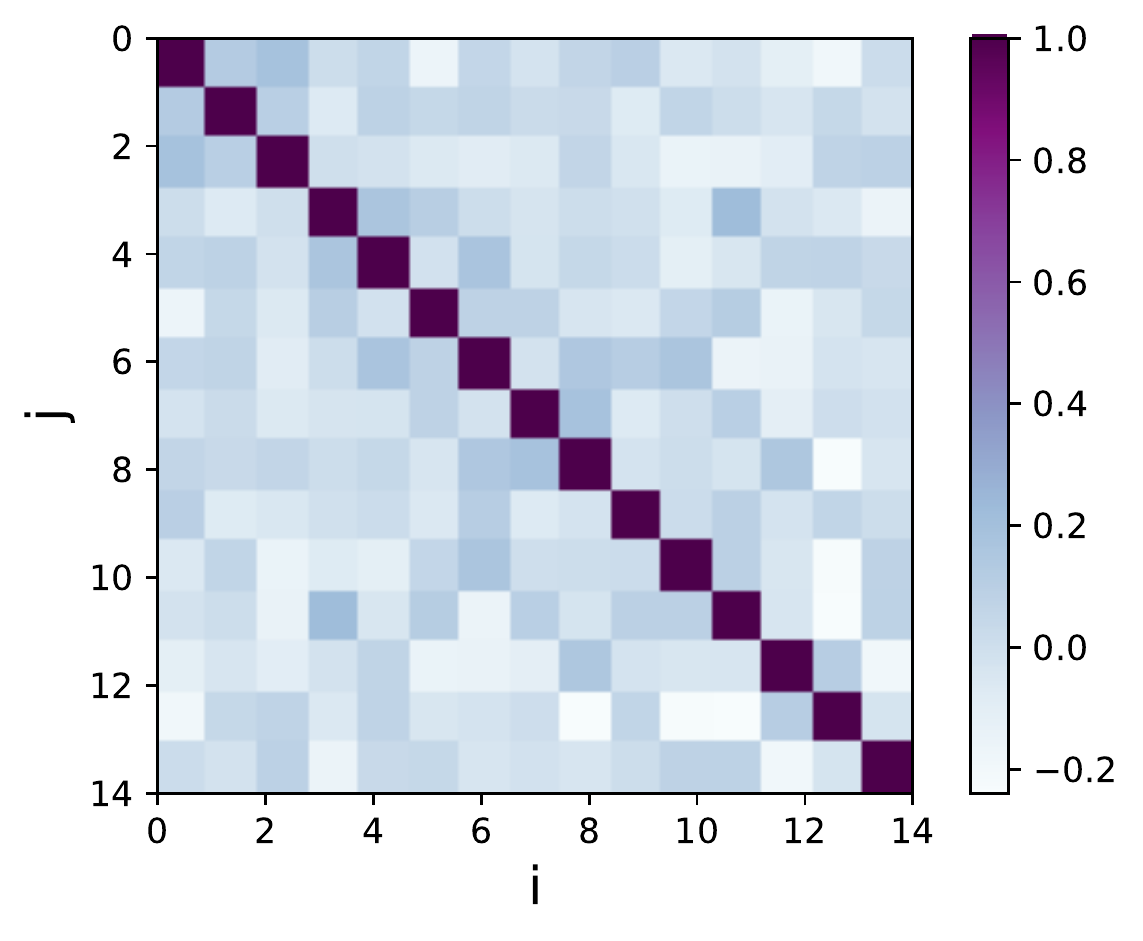}
    \caption{The normalized covariance matrix of the angular cross power spectrum (($\textrm{cov}[i, j] / \sqrt{\textrm{cov}[i, i] * \textrm{cov}[j, j]}$)), where $i$ and $j$ are labels of the bins.}
    \label{fig:cov}
\end{figure}

The off-diagonal elements of the covariance matrix are small compared to the on-diagonal elements (Fig.~\ref{fig:cov}), and the diagonal elements mostly agree with the theoretical values, calculated using Equation~\ref{eq:error}. In both the theoretically predicted error and the covariance matrix, the error in the cross power spectrum decreases with increasing $\ell$ for $\ell < 1200$. The shot noise of the quasars should be a constant contribution of the power spectrum error at all scales. On smaller scales, the error increases again, due to reconstruction noise in the lensing map.

The central value and the uncertainty of the bias estimate change slightly when we use the full covariance matrix, which gives a bias of $2.42 \pm 0.44$ with a significance of $5.4 \sigma$ and $\chi^2_{\textrm{th}} = 13.9$ for 14 degrees of freedom.

\section{Conclusions}
\label{sec:conclusions}
We studied the cross-correlation between the Planck CMB lensing convergence map and the eBOSS DR14 quasar map at redshift range $0.9 < z < 2.2$, with an effective redshift of $z_{\textrm{eff}} \approx 1.51$, and measure the quasar bias. We found correlation between CMB lensing and the eBOSS quasars, and a quasar bias $b_q = 2.43 \pm 0.45$ at $5.4\sigma$ significance, using the theoretically calculated covariance matrix. This is consistent with the result in \citet{Laurent2017}. We obtained the covariance matrix from the quasar mocks and lensing simulations, and found it consistent with the theoretical covariance matrix. While the theory curve is a good fit on most of the scales, the first bin shows low cross-correlation between CMB lensing and quasar clustering. The origin of this deficit of power at low-$\ell$ is not known at present.

We performed a simple null test for the cross power spectrum, and the result is mostly consistent with null, with the exception of two low-$\ell$ bins and one near $\ell_{\textrm{max}}$. We checked for several systematics and found no significant contributions from the considered contaminants.

Using the \citet{Tinker2010} model of the relation between halo mass function and clustering, we calculate a characteristic host halo mass for the eBOSS DR14 quasar catalog: $\log_{10}\left( \frac{M_{200}}{1 h^{-1} M_\odot} \right) = 12.54^{+0.25}_{-0.36}$. This is consistent with previous estimates of the quasar host halo mass at similar redshifts \citep{White2012, Laurent2017}. We also attempted to fit for a scale dependent bias, but did not find evidence for a scale dependent term. 

The significance and accuracy of the quasar bias measurement depend on the sample size and number density of the quasar survey \citep{Seljak2009}, so we would expect the signal-to-noise ratio of the detection using this method to improve, as eBOSS continues to expand its sample size \citep{Dawson2016} and new surveys such as DESI \citep{DESI2016} and Euclid \citep{Euclid2011} become operational. This will provide more precise measurements of the quasar bias as a function of redshift, scale, etc, and open paths to better understanding of the various properties of quasars, including the host halo mass and duty cycle \citep{Martini2001}. The improved bias measurement could also provide good constraints on the galaxy formation models \citep{Contreras2013}, general relativity and modified gravity \citep{Acquaviva2008}, and the properties of dark matter and dark energy \citep{Das2009}.

\section*{Acknowledgements}
We thank Siyu He, Anthony Pullen, Emmanuel Schaan, Blake Sherwin, Jeremy Tinker and Michael Wilson for helpful discussions. J.H. would like to thank Stephen Ebert and Pavel Motloch for helpful comments on the paper.
This work is based on observations made by the Planck satellite and the Apache Point Observatory. The Planck project (\url{http://www.esa.int/Planck}) is funded by the member states of ESA, and NASA. The SDSS-IV project (\url{http://www.sdss.org/}) is funded by the participating institutions, the National Science Foundation, the United States Department of Energy, and the Alfred P. Sloan Foundation. S.F. is funded by a Miller Fellowship at the University of California, Berkeley. 
S.H. thanks NASA for their support in grant number: NASA grant 15-WFIRST15-0008 and NASA ROSES grant 12-EUCLID12-0004. E.G. is supported by NSF grant AST1412966 and by the Simons Foundation through the Flatiron Institute. 

\bsp	
\label{lastpage}

\begin{thebibliography}{99}
\bibitem[\protect\citeauthoryear{Abazajian et al.}{2016}]{Abazajian2016}
Abazajian K. N., et al. 2016, preprint, (\href{https://arxiv.org/abs/1610.02743}{arXiv:1610.02743})
\bibitem[\protect\citeauthoryear{Acquaviva et al.}{2008}]{Acquaviva2008}
Acquaviva V., Hajian A., Spergel D. N., Das S. 2008, \href{https://journals.aps.org/prd/abstract/10.1103/PhysRevD.78.043514}{Phys. Rev. D, 78, 043514}
\bibitem[\protect\citeauthoryear{Aghanim et al.}{2008}]{Aghanim2008}
Aghanim N., Majumdar S., \& Silk J. 2008, Reports on Progress in Physics, 71, 066902. \href{https://arxiv.org/abs/0711.0518}{(arXiv:0711.0518)}
\bibitem[\protect\citeauthoryear{Amendola et al.}{2017}]{Amendola2017}
Amendola L., Menegoni E., Di Porto C., Corsi M., Branchini E. 2017, \href{https://journals.aps.org/prd/abstract/10.1103/PhysRevD.95.023505}{Phys. Rev. D 95, 023505}
\bibitem[\protect\citeauthoryear{Baron et al.}{2004}]{Baron2004}
Baron E., Nugent P. E., Branch D., Hauschildt P. H. 2004, \href{http://iopscience.iop.org/article/10.1086/426506/meta}{ApJ, 616, L91}
\bibitem[\protect\citeauthoryear{Blanton et al.}{2017}]{Blanton2017}
Blanton M. R., et al. 2017, \href{http://adsabs.harvard.edu/abs/2017arXiv170300052B}{AJ, 154, 28}
\bibitem[Bleem et al.(2012)]{2012ApJ...753L...9B} Bleem, L.~E., van Engelen, A., Holder, G.~P., et al.\ 2012, \apjl, 753, L9 

\bibitem[\protect\citeauthoryear{Bridle et al.}{2001}]{Bridle2001}
Bridle S. L., Zehavi I., Dekel A., Lahav O., Hobson M. P., Lasenby A. N. 2001, \href{https://academic.oup.com/mnras/article/321/2/333/980471}{MNRAS, 321, 333}
\bibitem[\protect\citeauthoryear{Cabr\'e et al.}{2008}]{Cabre2008}
Cabr\'e A., Fosalba P., Gazta\~naga E., Manera M. 2008, \href{https://academic.oup.com/mnras/article/381/4/1347/971484}{MNRAS, 381, 11}
\bibitem[\protect\citeauthoryear{Contreras et al.}{2013}]{Contreras2013}
Contreras S., Baugh C. M., Norberg P., Padilla N. 2013, \href{https://academic.oup.com/mnras/article/432/4/2717/996105}{MNRAS, 432, 2717}
\bibitem[\protect\citeauthoryear{Chuang et al.}{2014}]{Chuang2014}
Chuang C-H, Kitaura F-S, Prada F., Zhao C., Yepes G. 2014, \href{https://academic.oup.com/mnras/article/446/3/2621/2892948}{MNRAS, 446, 2621}
\bibitem[\protect\citeauthoryear{Das \& Spergel}{2009}]{Das2009}
Das S., Spergel D. N. 2009, \href{https://journals.aps.org/prd/abstract/10.1103/PhysRevD.79.043509}{Phys. Rev. D, 79, 043509}
\bibitem[\protect\citeauthoryear{Dawson et al.}{2013}]{Dawson2013}
Dawson K. S., Schlegel D. J., Ahn C. P., et al. 2013, \href{http://adsabs.harvard.edu/abs/2013AJ....145...10D}{AJ, 145, 10}
\bibitem[\protect\citeauthoryear{Dawson et al.}{2016}]{Dawson2016}
Dawson K. S., Kneib J-P, Percival W. J., et al. 2016, \href{http://iopscience.iop.org/article/10.3847/0004-6256/151/2/44/meta}{AJ, 151, 44}
\bibitem[\protect\citeauthoryear{DESI Collaboration et al.}{2016}]{DESI2016}
DESI Collaboration et al. 2016, preprint, \href{https://arxiv.org/abs/1611.00036}{(arXiv:1611.00036)}.
\bibitem[\protect\citeauthoryear{Desjacques et al.}{2016}]{Desjacques2016}
Desjacques V., Jeong D., Schmidt F., 2016, \href{https://www.sciencedirect.com/science/article/pii/S0370157317304192?via\%3Dihub}{Physics Reports, 733, 1}. (\href{https://arxiv.org/abs/1611.09787}{
arXiv:1611.09787})

\bibitem[\protect\citeauthoryear{DiPompeo et al.}{2014}]{DiPompeo2014}
DiPompeo M. A., et al. 2015, MNRAS 446, 3492. (\href{https://arxiv.org/abs/1411.0527}{
arXiv:1411.0527})

\bibitem[Ferraro \& Hill(2018)]{2018PhRvD..97b3512F} Ferraro, S., \& Hill, J.~C.\ 2018, \prd, 97, 023512 

\bibitem[\protect\citeauthoryear{Gazta\~naga et al.}{2012}]{Gaztanaga2012}
Gazta\~naga E., et al., 2012, \href{https://academic.oup.com/mnras/article/422/4/2904/1048436}{MNRAS, 422, 2904}

\bibitem[\protect\citeauthoryear{Geach et al.}{2013}]{Geach2013}
Geach J. E., et al., 2013, ApJ Letters, 776, L41. (\href{https://arxiv.org/abs/1307.1706}{arXiv:1307.1706})

\bibitem[\protect\citeauthoryear{Giannantonio et al.}{2016}]{Giannantonio2016}
Giannantonio T., et al., 2016, \href{http://adsabs.harvard.edu/abs/2016MNRAS.456.3213G}{MNRAS, 456, 3213}

\bibitem[\protect\citeauthoryear{Giusarma et al.}{2018}]{Giusarma2018}
Giusarma E., et al., 2018, preprint. \href{https://arxiv.org/abs/1802.08694}{(arXiv:1802.08694)}

\bibitem[\protect\citeauthoryear{Hinshaw et al.}{2012}]{Hinshaw2012}
Hinshaw G., et al. 2012, \href{http://iopscience.iop.org/article/10.1088/0067-0049/208/2/19/meta}{ApJS, 208, 19}
\bibitem[\protect\citeauthoryear{Hirata et al.}{2004}]{Hirata2004}
Hirata C. M., Padmanabhan N., Seljak U., Schlegel D.,
Brinkmann J. 2004, \href{https://journals.aps.org/prd/abstract/10.1103/PhysRevD.70.103501}{Phys. Rev. D, 70, 103501}

\bibitem[\protect\citeauthoryear{Hirata et al.}{2008}]{Hirata2008}
Hirata C. M., Ho S., Padmanabhan N., Seljak U.,
Bahcall N. 2008, \href{https://journals.aps.org/prd/abstract/10.1103/PhysRevD.78.043520}{Phys. Rev. D, 78, 043520}
\bibitem[\protect\citeauthoryear{Ho et al.}{2012}]{Ho2012}
Ho S., Cuesta A., et al. 2012, \href{http://adsabs.harvard.edu/abs/2012ApJ...761...14H}{ApJ, 761, 14}
\bibitem[\protect\citeauthoryear{Hu \& Okamoto}{2002}]{Hu2002}
Hu W. \& Okamoto T. 2002, \href{http://iopscience.iop.org/article/10.1086/341110/meta}{ApJ, 574, 566}
\bibitem[\protect\citeauthoryear{Kaiser}{1984}]{Kaiser1984}
Kaiser, N. 1984, \href{http://adsabs.harvard.edu/cgi-bin/bib_query?1984ApJ...284L...9K}{ApJ, 284, L9}
\bibitem[\protect\citeauthoryear{Kirk et al.}{2015}]{Kirk2015}
Kirk, D., et al. 2015 \href{https://academic.oup.com/mnras/article-abstract/459/1/21/2608819?redirectedFrom=fulltext}{MNRAS, 459, 21}
\bibitem[\protect\citeauthoryear{Kofman et al.}{1993}]{Kofman1993}
Kofman, L. A., Gnedin, N. Y., \& Bahcall, N. A. 1993, \href{http://adsabs.harvard.edu/full/1993ApJ...413....1K}{ApJ, 413, 1}
\bibitem[\protect\citeauthoryear{Kormendy \& Richstone}{1995}]{Kormendy1995}
Kormendy J. \& Richstone D. 1995, \href{http://adsabs.harvard.edu/abs/1995ARA\%26A..33..581K}{ARA\&A, 33, 581}
\bibitem[\protect\citeauthoryear{Kuntz}{2015}]{Kuntz2015}
Kuntz A. 2015, \href{https://www.aanda.org/articles/aa/abs/2015/12/aa26940-15/aa26940-15.html}{A\&A, 584, A53}
\bibitem[\protect\citeauthoryear{Laureijs et al.}{2011}]{Euclid2011}
Laureijs R., et al. 2011, preprint. \href{https://arxiv.org/abs/1110.3193}{(arXiv:1110.3193)}
\bibitem[\protect\citeauthoryear{Laurent et al.}{2017}]{Laurent2017}
Laurent P., Eftekharzadeh S., Le Goff J.-M. et al. 2017, \href{http://iopscience.iop.org/article/10.1088/1475-7516/2017/07/017/pdf}{JCAP, 07, 017}
\bibitem[\protect\citeauthoryear{Lewis et al.}{2000}]{Lewis2000}
Lewis A., Challinor A., \& Lasenby, A. 2000, \href{http://adsabs.harvard.edu/abs/2000ApJ...538..473L}{ApJ, 538, 473}
\bibitem[\protect\citeauthoryear{Lewis \& Challinor}{2006}]{Lewis2006}
Lewis A. \& Challinor A. 2006, Phys.
Rept., 429, 1. \href{https://arxiv.org/abs/astro-ph/0601594}{arXiv:astro-ph/0601594}
\bibitem[\protect\citeauthoryear{Limber}{1954}]{Limber1954}
Limber D. N. 1954, \href{http://adsabs.harvard.edu/abs/1954ApJ...119..655L}{ApJ, 119, 655}
\bibitem[Madhavacheril \& Hill(2018)]{2018arXiv180208230M} Madhavacheril, M.~S., \& Hill, J.~C.\ 2018, arXiv:1802.08230 

\bibitem[\protect\citeauthoryear{Martini \& Weinberg}{2001}]{Martini2001}
Martini P. \& Weinberg D. H. 2001,  
\href{http://iopscience.iop.org/article/10.1086/318331/meta}{ApJ, 547, 12}
\bibitem[\protect\citeauthoryear{Marziani \& Sulentic}{2014}]{Marziani2014}
Marziani P., Sulentic J. 2014,  
\href{https://www.sciencedirect.com/science/article/pii/S0273117713006418}{Advances in Space Research, 54, 1331}
\bibitem[\protect\citeauthoryear{Modi et al.}{2017}]{Modi2017}
Modi C., Castorina E., Seljak U. 2017, \href{https://academic.oup.com/mnras/article-abstract/472/4/3959/4093078?redirectedFrom=fulltext}{MNRAS, 472, 3959}
\bibitem[\protect\citeauthoryear{Moessner et al.}{1997}]{Moessner1997}
Moessner R., Jain B., Villumsen J. V. 1997, MNRAS, 294, 291. (\href{https://arxiv.org/abs/astro-ph/9708271}{arXiv:astro-ph/9708271})
\bibitem[\protect\citeauthoryear{Mortlock}{2015}]{Mortlock2015}
Mortlock D. J. 2015, preprint. \href{https://arxiv.org/abs/1511.01107}{(arXiv:1511.01107)}
\bibitem[\protect\citeauthoryear{Myers et al.}{2015}]{Myers2015}
Myers A. D., Palanque-Delabrouille N., Prakash A., et al. 2015, ApJS, 221, 27. \href{https://arxiv.org/abs/1508.04472}{arXiv:1508.04472}

\bibitem[Osborne et al.(2014)]{2014JCAP...03..024O} Osborne, S.~J., Hanson, D., \& Dor{\'e}, O.\ 2014, \jcap, 3, 024 

\bibitem[\protect\citeauthoryear{Peiris \& Spergel}{2000}]{Peiris2000}
Peiris H. V. \& Spergel D. N. 2000, \href{http://iopscience.iop.org/article/10.1086/309373/meta}{ApJ, 540, 605}
\bibitem[\protect\citeauthoryear{Peacock \& Smith}{2000}]{Peacock2000}
Peacock J. A. \& Smith R. E. 2000, \href{https://academic.oup.com/mnras/article/318/4/1144/957260}{MNRAS, 318, 1144}
\bibitem[\protect\citeauthoryear{Planck Collaboration et al.}{2015a}]{Planck2015-overview}
Planck Collaboration et al. 2015a, preprint, (\href{https://arxiv.org/abs/1502.01582}{arXiv:1502.01582})
\bibitem[\protect\citeauthoryear{Planck Collaboration et al.}{2015b}]{Planck2015-params}
Planck Collaboration et al. 2015b, \href{https://www.aanda.org/articles/aa/abs/2016/10/aa25830-15/aa25830-15.html}{A\&A, 594, A13}
\bibitem[\protect\citeauthoryear{Planck Collaboration et al.}{2015c}]{Planck2015-lensing}
Planck Collaboration et al. 2015c, \href{https://www.aanda.org/articles/aa/abs/2016/10/aa25941-15/aa25941-15.html}{A\&A, 594, A15}
\bibitem[\protect\citeauthoryear{Planck Collaboration et al.}{2015d}]{Planck2015-ccs}
Planck Collaboration et al. 2015d, \href{https://www.aanda.org/articles/aa/abs/2016/10/aa26914-15/aa26914-15.html}{A\&A, 594, A26}
\bibitem[\protect\citeauthoryear{Planck Collaboration et al.}{2015e}]{Planck2015-sz}
Planck Collaboration et al. 2015e, \href{https://www.aanda.org/articles/aa/abs/2016/10/aa25823-15/aa25823-15.html}{A\&A, 594, A27}
\bibitem[\protect\citeauthoryear{Planck Collaboration et al.}{2015f}]{Planck2015-gcc}
Planck Collaboration et al. 2015f, \href{https://www.aanda.org/articles/aa/abs/2016/10/aa25819-15/aa25819-15.html}{A\&A, 594, A28}
\bibitem[\protect\citeauthoryear{Planck Collaboration et al.}{2018}]{Planck2018-params}
Planck Collaboration et al. 2018, preprint, \href{https://arxiv.org/abs/1807.06209}{(arXiv:1807.06209)}
\bibitem[\protect\citeauthoryear{Pullen et al.}{2016}]{Pullen2016}
Pullen A. R., Alam S., He S., Ho S. 2016, \href{https://academic.oup.com/mnras/article-abstract/460/4/4098/2609153?redirectedFrom=fulltext}{MNRAS, 460, 4098}
\bibitem[\protect\citeauthoryear{Ross et al.}{2011}]{Ross2011}
Ross A. J., Ho S., et al., 2011, \href{https://academic.oup.com/mnras/article/417/2/1350/984555}{MNRAS, 417, 1350}
\bibitem[\protect\citeauthoryear{Sachs \& Wolfe}{1967}]{Sachs1967}
Sachs, R. K. \& Wolfe, A. M. 1967, \href{http://adsabs.harvard.edu/doi/10.1086/148982}{ApJ, 147, 73}
\bibitem[\protect\citeauthoryear{Salpeter}{1964}]{Salpeter1964}
Salpeter E. E. 1964, \href{http://adsabs.harvard.edu/abs/1964ApJ...140..796S}{ApJ, 140, 796}

\bibitem[Schaan \& Ferraro(2018)]{2018arXiv180406403S} Schaan, E., \& Ferraro, S.\ 2018, arXiv:1804.06403 

\bibitem[\protect\citeauthoryear{Schlegel et al.}{1998}]{Schlegel1998}
Schlegel D. J., Finkbeiner D. P., Davis M. 1998, \href{http://adsabs.harvard.edu/abs/1998ApJ...500..525S}{ApJ, 500, 525}
\bibitem[\protect\citeauthoryear{Scranton et al.}{2005}]{Scranton2005}
Scranton R., et al. 2005, \href{http://adsabs.harvard.edu/abs/2005ApJ...633..589S}{ApJ, 633, 589}
\bibitem[\protect\citeauthoryear{Seljak}{2000}]{Seljak2000}
Seljak U. 2000, \href{https://academic.oup.com/mnras/article/318/1/203/1143330}{MNRAS, 318, 203}
\bibitem[\protect\citeauthoryear{Seljak et al.}{2009}]{Seljak2009}
Seljak U., Hamaus N., Desjacques V. 2009, \href{https://journals.aps.org/prl/abstract/10.1103/PhysRevLett.103.091303}{Phys. Rev. Lett., 103, 091303}
\bibitem[\protect\citeauthoryear{Shen et al.}{2009}]{Shen2009}
Shen Y., Strauss M. A., Ross N. P., et al. 2009, \href{http://adsabs.harvard.edu/abs/2009ApJ...697.1656S}{ApJ, 697, 1656}
\bibitem[\protect\citeauthoryear{Sherwin et al.}{2012}]{Sherwin2012}
Sherwin B. D., Das S., et al. 2012, \href{https://journals.aps.org/prd/abstract/10.1103/PhysRevD.86.083006}{Phys. Rev. D 86, 083006}

\bibitem[\protect\citeauthoryear{Smith et al.}{2007}]{Smith2007}
Smith K. M., Zahn O., Dore O. 2007, Phys. Rev. D 76, 043510. (\href{https://arxiv.org/abs/0705.3980}{arXiv:0705.3980})

\bibitem[\protect\citeauthoryear{Smith et al.}{2003}]{Smith2003}
Smith R. E., Peacock J. A., Jenkins A., et al. 2003, \href{http://adsabs.harvard.edu/abs/2003MNRAS.341.1311S}{MNRAS, 341, 1311}
\bibitem[\protect\citeauthoryear{Spergel et al.}{1997}]{Spergel1997}
Spergel D. N., Bolte M., Freedman W. 1997, \href{http://www.pnas.org/content/94/13/6579}{PNAS, 94, 6579}

\bibitem[\protect\citeauthoryear{Story et al.}{2015}]{Story2015}
Story K. T., et al. 2015, ApJ, 810, 50. (\href{https://arxiv.org/abs/1412.4760}{arXiv:1412.4760})

\bibitem[\protect\citeauthoryear{Sunyaev \& Zel'dovich}{1980}]{Sunyaev1980}
Sunyaev R. A., \& Zel'dovich Ya. B. 1980, \href{https://www.annualreviews.org/doi/10.1146/annurev.aa.18.090180.002541}{ARA\&A, 18, 537}
\bibitem[\protect\citeauthoryear{Takahashi et al.}{2012}]{Takahashi2012}
Takahashi R., Sato M., Nishimichi T., Taruya A., Oguri M. 2012, \href{http://iopscience.iop.org/article/10.1088/0004-637X/761/2/152/meta}{ApJ, 761, 152}
\bibitem[\protect\citeauthoryear{Tinker et al.}{2010}]{Tinker2010}
Tinker J. L., Robertson B. E., Kravtsov A. V., et al. 2010, \href{http://iopscience.iop.org/article/10.1088/0004-637X/724/2/878/pdf}{ApJ, 724, 878}

\bibitem[van Engelen et al.(2014)]{2014ApJ...786...13V} van Engelen, A., Bhattacharya, S., Sehgal, N., et al.\ 2014, \apj, 786, 13 

\bibitem[\protect\citeauthoryear{Wandelt et al.}{2000}]{Wandelt2000}
Wandelt B. D., Hivon E., Gorski K. M. 2000, preprint. \href{https://arxiv.org/pdf/astro-ph/0008111.pdf}{(arXiv:astro-ph/0008111)}
\bibitem[\protect\citeauthoryear{Weinberg et al.}{2003}]{Weinberg2003}
Weinberg D. H., Dav'e R., Katz N., Kollmeier J. A. 2003, \href{https://aip.scitation.org/doi/abs/10.1063/1.1581786}{AIP Conference Proceedings 666, 157}
\bibitem[\protect\citeauthoryear{White \& Rees}{1978}]{White1978}
White S. D. M. \& Rees M. J. 1978, \href{http://adsabs.harvard.edu/abs/1978MNRAS.183..341W}{MNRAS, 183, 341}
\bibitem[\protect\citeauthoryear{White et al.}{2012}]{White2012}
White M., Myers A. D., Ross N. P., et al. 2012, \href{https://academic.oup.com/mnras/article/424/2/933/1005224}{MNRAS, 424, 933}
\bibitem[\protect\citeauthoryear{Zaldarriaga \& Seljak}{1999}]{Zaldarriaga1999}
Zaldarriaga M. \& Seljak U. 1999, \href{https://journals.aps.org/prd/abstract/10.1103/PhysRevD.59.123507}{Phys. Rev. D 59, 123507}
\bibitem[\protect\citeauthoryear{Zhao et al.}{2016}]{Zhao2016}
Zhao G.-B., Wang Y., Ross A. J., et al. 2016, \href{https://academic.oup.com/mnras/article-abstract/457/3/2377/2588920?redirectedFrom=fulltext}{MNRAS, 457, 2377}
\end{thebibliography}
\end{document}